\documentclass{llncs}
\usepackage{wrapfig}
\usepackage{llncsdoc}
\usepackage{epsfig,psfrag}
\usepackage{latexsym,amssymb,amsmath}
\usepackage{longtable}
\usepackage{tikz,pgf}
\usepackage{subfig}
\usepackage{wrapfig}
\usepackage{rotating}

\usepackage{graphicx}
\usepackage{epstopdf}
\usepackage{color}

\usepackage{amsmath,amsfonts}
\newcommand{\bqn}{\begin{eqnarray}}
\newcommand{\eqn}{\end{eqnarray}}
\newcommand{\bq}{\begin{eqnarray*}}
\newcommand{\eq}{\end{eqnarray*}}

\usepackage{url}
\usepackage{hyperref}
\hypersetup{colorlinks,%
citecolor=blue,%
filecolor=blue,%
linkcolor=red,%
urlcolor=blue,%
pdftex}

\newcommand{\red}[1]{\textcolor{red}{#1}}

\begin{document}

\title{Exact Combinatorial Inference for Brain Images}

\author{Moo K. Chung$^1$, Zhan Luo$^1$, Alex D. Leow$^2$, 
Andrew L. Alexander$^1$, 
\\Richard J. Davidson$^1$, H. Hill Goldsmith$^1$}

\institute{
University of Wisconsin-Madison, $^2$University of Illinois-Chicago, 
USA\\
\vspace{0.3cm}
\red{\tt mkchung@wisc.edu}}

\maketitle


\begin{abstract}

The permutation test is known as the exact test procedure in statistics. However, often it is not exact in practice and only an approximate method since only a small fraction of every possible permutation is generated. Even for a small sample size, it often requires to generate tens of thousands permutations, which can be a serious computational bottleneck. In this paper, we propose a novel combinatorial inference procedure that enumerates all possible permutations combinatorially without any resampling. The proposed method is validated against the standard permutation test in simulation studies with the ground truth. The method is further applied in twin DTI study in determining the genetic contribution of the minimum spanning tree of the structural brain connectivity. 
\end{abstract}

\section{Introduction}
The permutation test  is perhaps the most widely used nonparametric test procedure in sciences \cite{chung.2017.IPMI,thompson.2001,zalesky.2010}. It is known as the exact test in statistics 
since the distribution of the test statistic under the null hypothesis can be exactly computed if we can calculate all possible values of the test statistic under every possible permutation. 
Unfortunately, generating every possible permutation  for whole images is still extremely time consuming even for modest sample size.

When the total number of permutations are too large, various resampling techniques have been proposed to speed up the computation in the past. In the resampling methods, only a small fraction of possible permutations are generated and the statistical significance is computed {\em approximately.} This approximate permutation test is the most widely used version of the permutation test. In most of brain imaging studies, 5000-1000000 permutations are often used, which puts the total number of generated permutations usually less than 1\% of all possible permutations. 
In \cite{thompson.2001}, for instance, 1 million permutations out of ${40 \choose 20}$ possible permutations (0.07\%) were generated using a super computer.  

In this paper, we propose a novel {\em combinatorial inference} procedure that enumerates all possible permutations combinatorially and simply avoids resampling that is slow and approximate. Unlike the permutation test that takes few hours to few days in a desktop, our exact procedure takes few seconds. Recently combinatorial approaches for statistical inference are emerging as an powerful alternative to existing statistical methods \cite{neykov.2016,chung.2017.IPMI}.  Neykov et al. proposed a combinatorial  technique for graphical models. However, their approach still relies on bootstrapping, which is another resampling technique and still approximate  \cite{neykov.2016}. Chung et al. proposed another combinatorial approach for brain networks but their method is limited to integer-valued graph features  
\cite{chung.2017.IPMI}. Our main contributions of this paper are as follows.

\begin{wrapfigure}{r}{0.45\textwidth}
\vspace{-1cm}
 \centering
\includegraphics[width=1\linewidth]{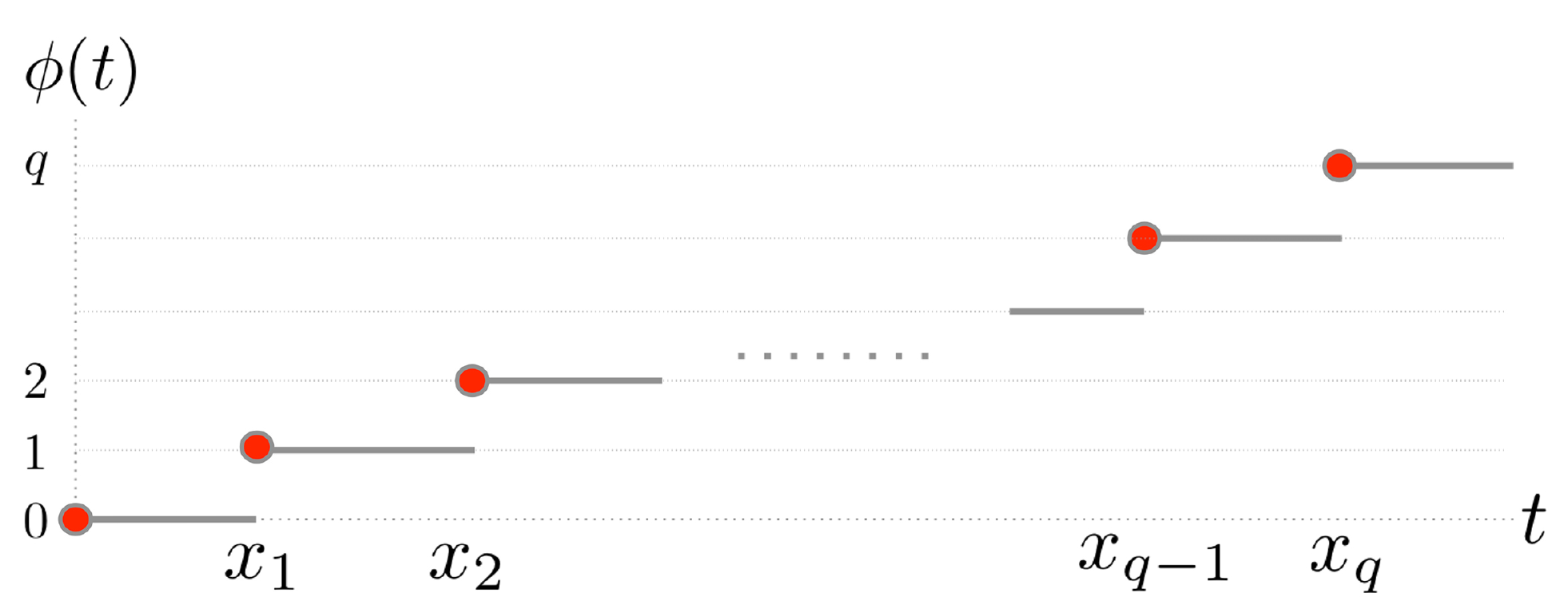}
\caption{\footnotesize Monotone sequence $x_1 < x_2 < \cdots < x_q$ is mapped to integers between 1 and $q$ {\em without} any gap via $\phi(t)$.}
\label{fig:permute-mono}
\vspace{-0.5cm}
\end{wrapfigure}

1) A new combinational approach for the permutation test that does not require 
resampling. While the permutation tests require exponential run time and approximate \cite{chung.2017.IPMI}, our combinatorial approach requires $\mathcal{O}(n^2)$ run time and exact. 2) Showing that the proposed method is a more sensitive and powerful alternative to the existing permutation test through extensive simulation studies. 3) A new formulation for testing the brain network differences by using  the minimum spanning tree differences. The proposed  
framework is applied to a twin DTI study in determining the heritability of the structural brain network.

\section{Exact combinatorial inference}
\label{sec:ETI}

The method in this paper extends our previous {\em exact topological inference} \cite{chung.2017.IPMI}, which is limited to integer-valued monotone functions from graphs. Through Theorem \ref{thm:monotone}, we extend the method to any arbitrary monotone function.

\begin{definition} For any  sets $G_1$ and $G_2$ satisfying $G_1\subset G_2$, function $f$ is strictly {\em monotone}  if it satisfies $f(G_1) <  f(G_2)$. $\subset$ denotes the strict subset relation.
\end{definition}

\begin{theorem}
\label{thm:monotone}
Let $f$ be a monotone function on the nested set $G_1\subset G_2\subset \cdots\subset G_q.$
Then there exists a nondecreasing function $\phi$ such that 
$\phi \circ f(G_j) = j$.
\end{theorem}
{\em Proof.} We prove the statement by actually constructing such a function. Function $\phi$ is constructed as follows. Let $ x_j = f(G_j)$. Then obviously $x_1 < x_2 < \cdots < x_q.$
Define an increasing step function $\phi$ such that
$$\phi(t) = 0  \mbox{ if } t < x_1,  \quad \phi(t) = j  \mbox { if } x_{j} \leq t < x_{j+1},  \quad  \phi(t) =  q \mbox{ if }x_q \leq t
.$$
The step function $\phi$ is illustrated in Figure \ref{fig:permute-mono}. Then it is straightforward to see that $\phi \circ f (G_j) = j$ for all $1 \leq j \leq q$. Further $\phi$ is nondecreasing. $\square$

Consider two nested sets $ F_1\subset F_2\subset \cdots\subset F_q$ and  $ G_1\subset G_2\subset \cdots\subset G_q. $
We are interested in testing the null hypothesis $H_0$ of the equivalence of two monotone functions defined on the nested sets:
\bq f(F_1) <  f(F_2) < \cdots < f(F_q)  \quad \mbox{ vs. }   \quad g(G_1) < g(G_2) < \cdots < g(G_q). \label{eq:FG} \eq
We have nondecreasing functions $\phi$ and $\psi$ on $f(F_j)$ and $g(G_j)$ respectively that satisfies the condition Theorem \ref{thm:monotone}. We use psudo-metric
$$D_q = \max_{t} \big| \phi (t) - \psi (t) \big|$$
as a test statistic that measures the similarity between two monotone functions. The use of maximum removes the problem of multiple comparisons. The distribution of $D_q$ can be determined  by combinatorially.

\begin{theorem}
\label{thm:lim1}
$P (D_q \geq d)   = 1 - \frac{A_{q,q}}{{2q \choose q}},$
where $A_{u,v}$ satisfies $A_{u,v} = A_{u-1,v} + A_{u, v-1}$
with the boundary condition $A_{0,v}=A_{u,0}=1$ within band $|u - v| < d$ and initial condition $A_{0,0} =0$ for $u,v \geq 1$.
\end{theorem} 
{\em Proof.}
Let $x_j = \phi \circ f(F_j)$ and $y_j = \phi \circ g(G_j)$. 
From Theorem \ref{thm:monotone},  the sequences $x_1, \cdots,  x_q$ and $y_1, \cdots,  y_q$ are monotone and integer-valued between 1 and $q$ {\em without} any gap. Perform the permutation test on the sorted sequences. If we identify each $x_j$ as moving one grid to right and $y_j$ as moving one grid to up, each permutation is mapped to a walk from $(0,0)$ to $(q,q)$. There are total ${2q \choose q}$ number of such paths and each permutation can be mapped to a walk uniquely.

\begin{wrapfigure}{r}{0.3\textwidth}
\vspace{-0.5cm}
 \centering
\includegraphics[width=1\linewidth]{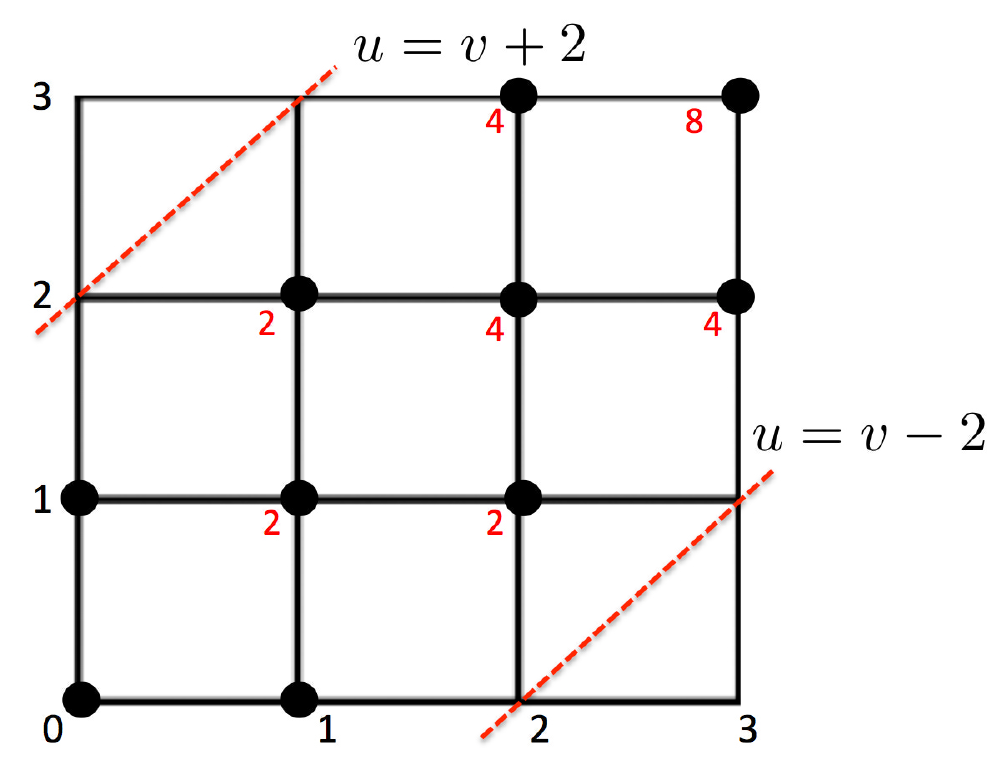} 
\caption{\footnotesize In this example, $A_{u,v}$ is computed within the boundary (dotted red line) from (0,0) to (3,3).}
\label{fig:App}
\vspace{-0.5cm}
\end{wrapfigure}

Note $\max_{1 \leq j \leq q} \big|x_j - y_j \big| < d$ if and only if
$ |x_j - y_j | < d \mbox{  for all } 1 \leq j \leq q. \label{eq:xy} $
Let $A_{q,q}$ be the total number of paths within $| x- y| < d$ (Fig. \ref{fig:App} for an illustration). Then it follows  that $A_{q,q}$ is iteratively given as
$A_{u,v} = A_{u-1,v} + A_{u, v-1}$ with $A_{0,0} =0, A_{0,v}=A_{u,0}=1, $ within $|u - v| < d$. Thus
$P(D_q< d) = \frac{A_{q,q}}{{2q \choose q}}$. $\square$

For example, $P(D_3 \geq 2)$ is computed sequentially as follows (Fig. \ref{fig:App}).  We start with the bottom left corner $A_{0,0} = 0$ and move right or up toward the upper corner. $A_{1,0} = 1, A_{0,1}=1 \to A_{1,1} = A_{1,0} + A_{0,1} \to \cdots \to A_{3,3} = A_{3,2} + A_{2,3} =8$. The probability is then $P(D_3 \geq 2) = 1- 8/ {6 \choose 3}=0.6$. The computational complexity of the combinatorial inference is $\mathcal{O}(q^2)$ for computing $A_{q,q}$ in the grid while the permutation test is exponential.

\section{Inference on minimum spanning trees}
As a specific example of how to apply the method, we show how to test for shape differences in minimum spanning trees (MST) of graphs. MST are often used in speeding up computation and simplifying complex graphs as simpler trees \cite{wu.1993}. We et al. used MST in edge-based segmentation of lesion in brain MRI \cite{wu.1993}. Stam et al.  used MST as an unbiased skeleton representation of complex brain networks \cite{stam.2014}. Existing statistical inference methods on MST rely on using graph theory features on MST such as the average path length. Since the probability distribution of such features are often not well known, the permutation test is frequently used, which is not necessarily exact or effective. Here, we apply the proposed combinatorial inference for testing the shape differences of MST.

For a graph with $p$ nodes, MST is often constructed using Kruskal's algorithm, which is a greedy algorithm with runtime $\mathcal{O}(p\log p)$. The algorithm  starts with an edge with the smallest weight. Then add an edge with the next smallest  weight. This sequential process continues while avoiding a loop and  generates a spanning tree with the smallest total edge weights. Thus, the edge weights in MST correspond to the order, in which the edges are added in the construction of MST.  Let $M_1$ and $M_2$ be the MST corresponding to $p \times p$ connectivity matrices $C_1$ and $C_2$. We are interested in testing hypotheses
$$H_0: M_1= M_2 \quad \mbox{ vs. } \quad H_1: M_1 \neq M_2.$$
The statistic for testing $H_0$ vs. $H_1$ is as follows. Since there are $p$ nodes, there are $p-1$ edge weights in MST. Let $ w_1^1 < w_2^1 < \cdots < w_{p-1}^1 \label{eq:NI-w1}$ and
$ w_1^2 < w_2^2 < \cdots < w_{p-1}^2 \label{eq:NI-w2} $
be the sorted edge weights in $M_1$ and $M_2$ respectively. These correspond to the order MST are constructed in Kruskal's algorithm. $w_j^1$ and $w_j^2$ are edge weights obtained in the $j$-th iteration of Kruskal's algorithm. Let $\phi$ and $\psi$ be monotone step functions that map the the edge weights obtained in the $j$-th iteration to 
integer $j$, i.e.,
$\phi(w_j^1) = j, \; \psi(w_j^2) =j.$ $\phi$ and $\psi$ can be interpreted as the number of nodes added into MST in the $j$-th iteration.

\begin{figure}[t!]
\centering
\includegraphics[width=1\linewidth]{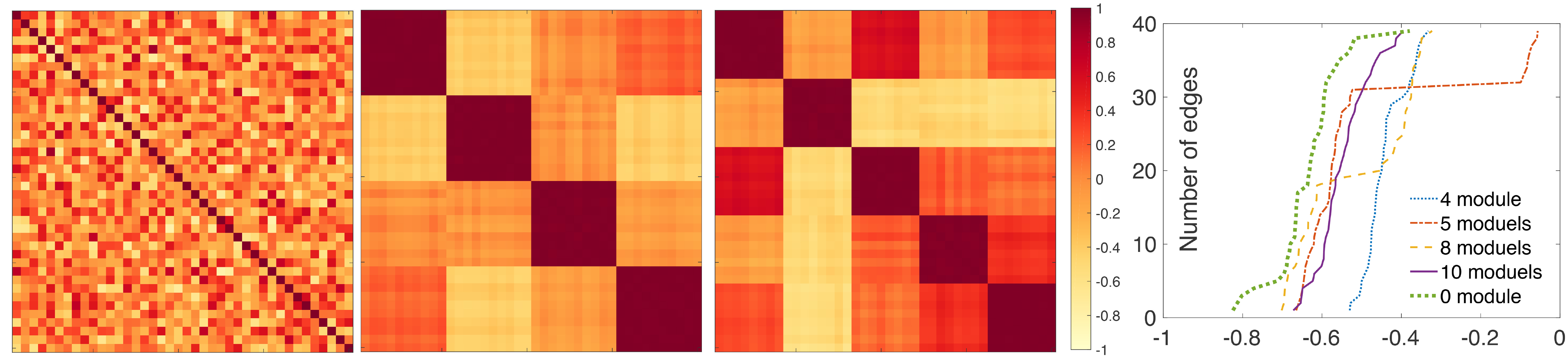}
\caption{\footnotesize Randomly simulated correlation matrices with 0, 4 and 5 modules. The plot shows the number of nodes over the largest edge weights added into MST construction during Kruskal's algorithm for 4, 5, 8, 10 and 0 modules.}
\label{fig:simul-modular2}
\end{figure}

\section{Validation and comparisons}

For validation and comparisons, we simulated the random graphs with the ground truth. We used $p=40$ nodes and $n=10$ images, which makes possible permutations to be exactly ${10 + 10 \choose 10}=184756$ making the permutation test manageable. The data matrix $X_{n \times p}=(x_{ij}) = ({\bf x}_1, {\bf x}_2, \cdots, {\bf x}_p)$  is simulated as standard normal in each component, i.e.,
$x_{ij} \sim N(0,1). \label{eq:simul-xij}$ or equivalently each column is multivariate normal
${\bf x}_j \sim N(0,I)$ with  identity matrix as the covariance.   

\begin{table}[t]
\caption{\footnotesize Simulation results given in terms of $p$-values. In the case of no network differences (0 vs. 0 and 4 vs. 4), higher $p$-values are better. In the case of network differences (4 vs. 5, 4 vs. 8 and 5 vs. 10), smaller $p$-values are better. \label{table:simulation}} 
\centering
\tabcolsep 4pt
\begin{tabular}{|c|c|c|c|c|c|c|} 
\hline
& Combinatorial  &    Permute 0.1\% & Permute 0.5\% & Permute 1\% \\
\hline
0 vs. 0&	 $0.831  \pm  0.187$ & $0.746   \pm 0.196$ & $0.745  \pm  0.195$& $ 0.744   \pm 0.196$\\
4 vs. 4&  $0.456  \pm  0.321$ & $0.958   \pm 0.075$ & $0.958  \pm  0.073$ & $ 0.958  \pm  0.073$ \\
4  vs. 5 &  $0.038  \pm  0.126$  & $0.381   \pm 0.311$ &$0.377    \pm 0.311$&$ 0.378   \pm 0.311$ \\
4  vs. 8 &  $0.053  \pm  0.138$  & $ 0.410 \pm   0.309$ &$ 0.411  \pm  0.306$&$ 0.411  \pm  0.306$ \\
5 vs. 10 & $ 0.060  \pm 0.126$ &  $    0.391  \pm  0.283$ & $0.395    \pm 0.284$&  $0.395   \pm 0.283  $\\ 
\hline
\end{tabular}
\end{table}

Let $Y=(y_{ij}) = ({\bf y}_1, \cdots, {\bf y}_p) = X$. So far, there is no statistical dependency  between nodes in $Y$. We add the following block modular structure to $Y$. We assume there are $k=4,5,8,10, 40$ modules and each module consists of $c=p/k = 10,8,5,4,1$ number of nodes. Then for the $i$-th node  in the $j$-th module, we simulate
\bqn {\bf y}_{c(j-1)+i}= {\bf x}_{c(j-1)+1} + N(0,\sigma I) \; \quad \mbox{ for } 1 \leq i \leq c,  1 \leq j \leq k \label{eq:yxj} \eqn
with $\sigma=0.1$.
Subsequently, the connectivity matrix $C=(c_{ij})$ is given by $c_{ij} = corr({\bf y}_i, {\bf y}_j)$. This introduces the block modular structure in the correlation network (Fig. \ref{fig:simul-modular2}). For 40 modules, each module consists of just 1 node, which is basically a network with 0 module. 

Using (\ref{eq:yxj}), we simulated random networks with 4,5,8, 10 and 0 modules. For each network, we obtained MST and  computed the distance $D$ between networks. We computed the $p$-value using the combinatorial method. In comparison, we performed the permutation tests by permuting the group labels and generating 0.1, 0.5 and 1\% of every possible permutation. The procedures are repeated 100 times and the average results are reported in Table \ref{table:simulation}.

In the case of no network differences (0 vs. 0 and 4 vs. 4), higher $p$-values are better. The combinatorial method and the permutation tests all performed well for no network difference. In the case of network differences (4 vs. 5, 4 vs. 8 and 5 vs. 10), smaller $p$-values are better. The combinatorial method performed far superior to the permutation tests. None of the permutation tests detected modular structure differences. The proposed combinatorial approach on MST seems to be far more sensitive in detecting modular structures.  The performance of the permutation test does not improve even when we sample 10\% of all possible permutations. The permutation test doesn't converge rapidly with increased samples. The codes for performing exact combinatorial inference as well as simulations can be obtained from \url{http://www.stat.wisc.edu/~mchung/twins}.

\begin{figure}[t]
\centering
\includegraphics[width=1\linewidth]{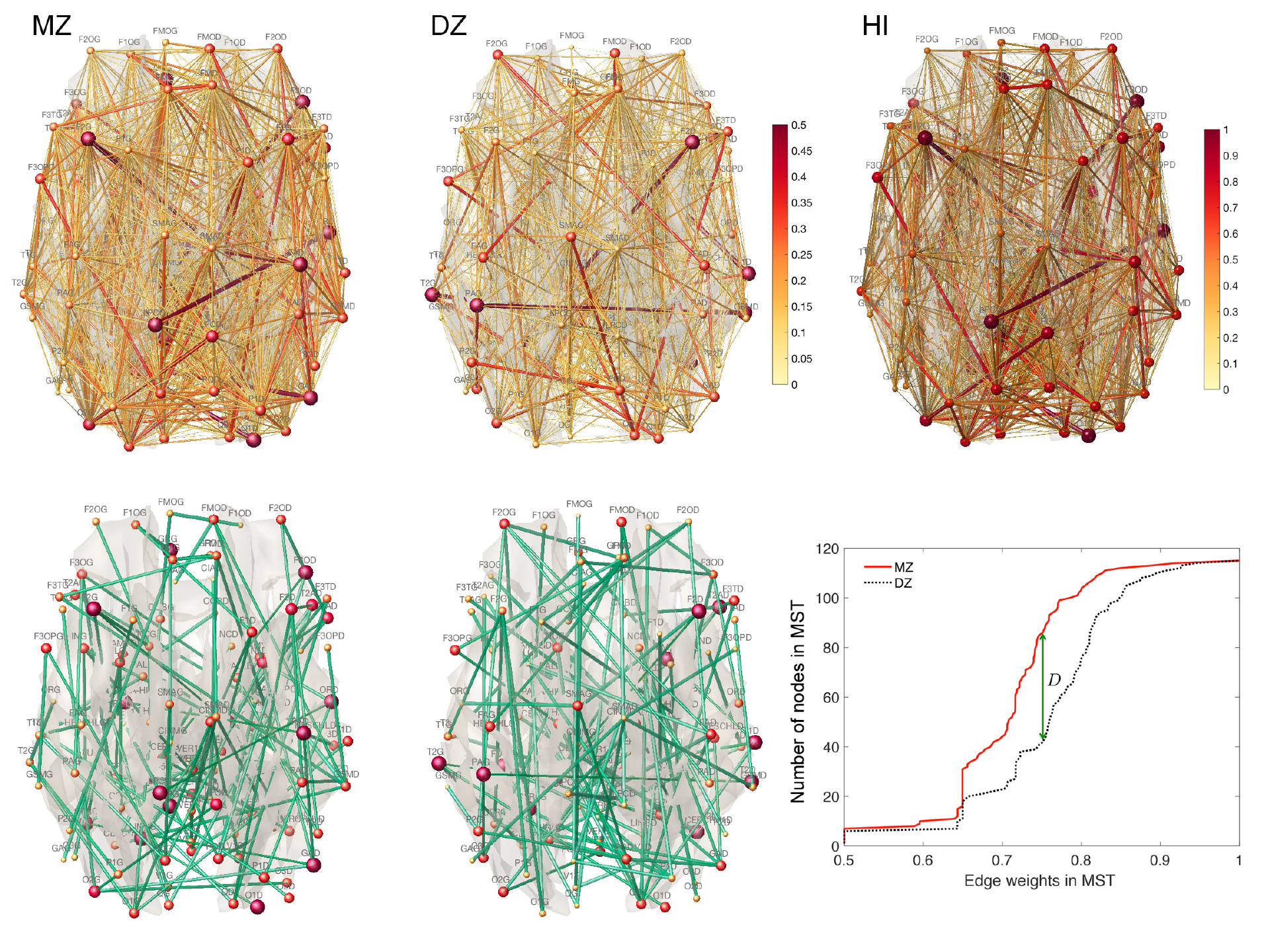}
\caption{\footnotesize Top: Correlation network of MZ- and DZ-twins and heritability index (HI). 
Bottom:  Minimum spanning trees (MST) constructed using Kruskal's algorithm on 1-correlation. Plot: The  number of added nodes is plotted over the largest edge weights of MST for MZ- (solid red) and DZ-twins (dotted black) during the MST construction. The pseudo-metric $D$ is  46 at edge weight 0.75 (corresponding to correlation 0.25).}
\label{fig:MSTtwins}
\end{figure}

\section{Application to twin DTI study}

{\em Subjects.} The method is applied to 111 twin pairs of diffusion weighted images (DWI).
Participants were part of 
the Wisconsin Twin Project \cite{goldsmith.2007}. 
58 monozygotic (MZ) and 53 same-sex dizygotic (DZ) twins 
were used in the analysis. We are interested in knowing the extent of the genetic influence on the structural brain network of these participants and determining its statistical significance between MZ- and DZ-twins. 
Twins were scanned in a 3.0 Tesla GE Discovery MR750 scanner with a 32-channel receive-only head coil. 
Diffusion tensor imaging (DTI) was performed using a three-shell diffusion-weighted, spin-echo, echo-planar imaging  sequence. A total of 6 non-DWI (b=0 s$\cdot$mm2) and 63 DWI with non-collinear diffusion encoding directions were collected at b=500, 800, 2000 (9, 18, 36 directions). Other  parameters were TR/TE = 8575/76.6 ms; parallel imaging; 
 flip angle = 90$^\circ$; isotropic 2mm resolution (128$\times$128 matrix with 256 mm field-of-view). 

{\em Image processing.}  FSL were used to correct for eddy current related distortions, head motion and field inhomogeneity. 
Estimation of the diffusion tensors at each voxel was performed using non-linear tensor estimation in CAMINO. 
DTI-TK was used for constructing the study-specific template \cite{hanson.2013}. Spatial normalization was performed for tensor-based white matter alignment using a non-parametric diffeomorphic registration method \cite{zhang.2006}. 
Each subject's tractography was constructed in the study template space 
using the streamline method, and tracts were terminated at FA-value less than 0.2 and deflection angle greater than 60 degree \cite{lazar.2003.HBM}. 
Anatomical Automatic Labeling (AAL) with $p=116$ parcellations were used to construct $p \times p$ structural connectivity matrix that counts the number of white matter fiber tracts between the parcellations \cite{tzourio.2002}.

{\em Exact combinatorial inference.} From the individual structural connectivity matrices, we computed pairwise twin correlations in each group using  Spearman's correlation. The resulting twin  correlations matrices $C_{MZ}$ and $C_{DZ}$ (edge weights in Fig. \ref{fig:MSTtwins}) were used to compute the heritability index (HI) through  Falconer's formula, which determines the amount of variation due to genetic influence in a population:
$\mbox{HI} = 2 ( C_{MZ} - C_{DZ} )$ \cite{chung.2017.IPMI}.
Although HI provides quantitative measure of heritability, it is not clear if it is statistically significant. We tested the significance of HI by testing the equality of $C_{MZ}$ and $C_{DZ}$. We used $1-C_{MZ}$ and $1-C_{DZ}$ as edge weights in finding MST using Kruskal's algorithm. This is equivalent to using $C_{MZ}$ and $C_{DZ}$ as edge weights in finding the {\em maximum spanning trees}. Fig. \ref{fig:MSTtwins} plot shows how the number of nodes increase as the edges are added into the MST construction. At the same edge weights, MZ-twins are more connected than DZ-twins in MST. This implies MZ-twins are connected less in lower correlations and connected more in higher correlations. 

\begin{wrapfigure}{rt}{0.5\textwidth}
\vspace{-1cm}
\centering
\includegraphics[width=1\linewidth]{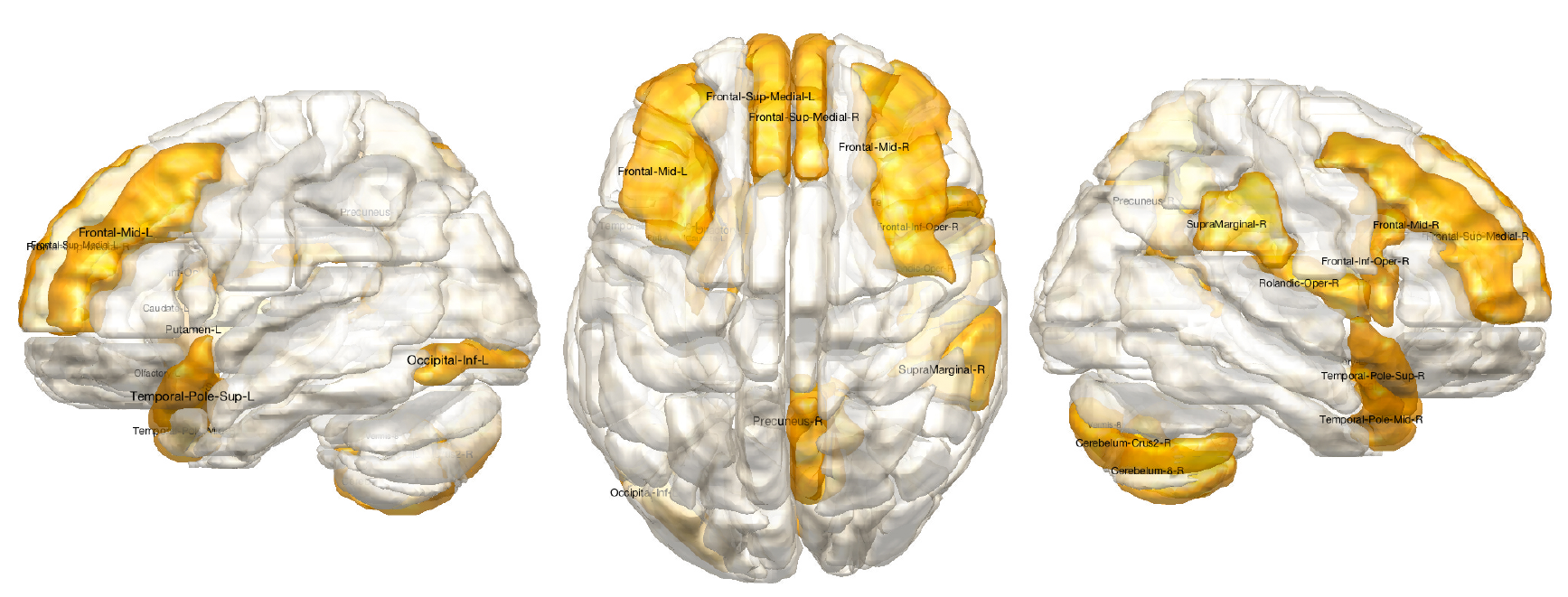}
\caption{\footnotesize Regions corresponding to the maximum difference between MST of MZ- and DZ-twins.}
\label{fig:AAL-regions}
\vspace{-0.5cm}
\end{wrapfigure}

{\em Results.} At edge weight 0.75, which is the maximum gap and corresponding to correlation 0.25, the observed distance $D$ was 46. The corresponding $p$-value was computed as 
$P(D \geq 46) = 1.57  \times 10^{-8}$. 
The localized regions of brain that genetically contribute the most can also be identified by identifying the nodes of connections around edge weight 0.75 ($0.75 \pm 0.2$). The following AAL regions are identified as the region of statistically significant MST differences: 
Frontal-Mid-L,
    Frontal-Mid-R,
    Frontal-Inf-Oper-R,
    Rolandic-Oper-R,
    Olfactory-L,
    Frontal-Sup-Medial-L,
    Frontal-Sup-Medial-R,
    Occipital-Inf-L,
    SupraMarginal-R,
    Precuneus-R,
    Caudate-L,
    Putamen-L,
    Temporal-Pole-Sup-L,
    Temporal-Pole-Sup-R,
    Temporal-Pole-Mid-R,
    Cerebelum-Crus2-R,
    Cerebelum-8-R,
    Vermis-8 (Fig. \ref{fig:AAL-regions}). The identified frontal and temporal regions are overlapping with the previous MRI-based twin study \cite{thompson.2001}.

\section{Conclusion and discussion}

We presented the novel exact combinatorial inference method that outperforms the traditional permutation test. The main innovation of the method is that it works for any monotone features.  Given any two sets of measurements, all it requires is to sort them and we can apply the method. Thus, the method can be applied to wide variety of applications. For this study, we have shown how to apply  the method in testing the shape differences in MST of the structural brain networks. The method was further utilized in localizing brain regions influencing such differences. 

In graphs, there are many monotone functions including the number of connected components, total node degrees and the sorted eigenvalues of  graph Laplacians. These monotone functions can all be used in the proposed combinatorial inference. The proposed method is also equally applicable to wide variety of monotonically decreasing graph features such as the largest connected components \cite{chung.2017.IPMI}. If $\phi \circ f$ is monotonically decreasing, $ - \phi \circ f$ is monotonically increasing, thus the same method is applicable to decreasing functions. The applications of other features are left for future studies.

\section*{Acknowledgements}
This work was supported by NIH grants R01 EB022856, R01 MH101504,  P30 HD003352, U54 HD09025, UL1 TR002373. We thank Nagesh Adluru, Yuan Wang, Andrey Gritsenko (University of Wisconsin-Madison), Hyekyoung Lee (Seoul National University), Zachery Morrissey (University of Illinois-Chicago) and Sourabh Palande, Bei Wang (University of Utah) for valuable supports. 
\bibliographystyle{plain}
\bibliography{reference.2018.05.27}

\end{document}